\documentclass[9pt,twocolumn,twoside]{pnas-new}
\newcommand\mymapsto{\mathrel{\ooalign{$\rightarrow$\cr%
  \kern-.15ex\raise.275ex\hbox{\scalebox{1}[0.522]{$\mid$}}\cr}}}
\DeclareMathOperator{\Tr}{Tr}
\usepackage{stmaryrd}

 \templatetype{pnasresearcharticle} 

\title{Zoo Guide to Network Embedding}

\author[1,2,*]{Anthony Baptista}
\author[2,3,4]{Rubén J. Sánchez-García}
\author[5,6,7]{Anaïs Baudot}
\author[1,2,*]{Ginestra Bianconi}

\affil[1]{School of Mathematical Sciences, Queen Mary University of London, London, E1 4NS, United Kingdom}
\affil[2]{The Alan Turing Institute, The British Library, London, NW1 2DB, United Kingdom}
\affil[3]{School of Mathematical Sciences, University of Southampton, Southampton SO17 1BJ, United Kingdom}
\affil[4]{Institute for Life Sciences, University of Southampton, Southampton SO17 1BJ, United Kingdom}
\affil[5]{Aix-Marseille Univ, INSERM, MMG, Marseille, France}
\affil[6]{Barcelona Supercomputing Center, Barcelona, Spain}
\affil[7]{CNRS, Marseille, France}

\leadauthor{A.Baptista} 

\authorcontributions{A.Bap., G. B. and  A.Bau. designed the research; A.Bap. performed the research; A.Bap. R.S-G,  G.B and A.Bau. wrote the paper.}
\authordeclaration{The authors declare no conflict of interest.}
\correspondingauthor{\textsuperscript{*}To whom correspondence should be addressed. E-mail: anthony.baptista@qmul.ac.uk, ginestra.bianconi@gmail.com}


\begin{abstract}
Networks have provided extremely successful models of data and complex systems. Yet, as combinatorial objects, networks do not have in general intrinsic coordinates and do not typically lie in an ambient space. The process of assigning an embedding space to a network  has attracted lots of interest in the past few decades, and has been efficiently applied to fundamental problems in network inference, such as link prediction, node classification, and community detection. In this review, we  provide a user-friendly guide to the network embedding literature and current trends in this field which will allow the reader to navigate through the complex landscape of methods and approaches emerging from the vibrant research activity on these subjects. 
\end{abstract}

\dates{This manuscript was compiled on \today}

\begin{document}

\maketitle
\thispagestyle{firststyle}
\ifthenelse{\boolean{shortarticle}}{\ifthenelse{\boolean{singlecolumn}}{\abscontentformatted}{\abscontent}}{}

\dropcap{N}etworks are simple yet powerful and versatile models to represent and analyse complex data and systems across a wide variety of user domains and research fields \cite{newman2018networks, barabasi2013network}. In social sciences, social networks are useful for different tasks, such as items or friends classifications, friends recommendations or targeted advertising \cite{borgatti2018analyzing}. Using social networks, community detection or link prediction can help to better understand the spreading process of rumours or epidemics \cite{pastor2015epidemic}. In biology, link prediction in biological networks is commonly used for predicting new interactions between proteins, new therapeutic applications for existing drugs or new gene-disease associations \cite{junker2011analysis, alm2003biological}. Overall, the study of networks  as mathematical models has developed into the fully established discipline of Network Science \cite{newman2018networks, barabasi2013network}. 

Networks are intrinsically combinatorial objects (i.e., interconnected nodes, where certain pairs of nodes are connected by links), with no \textit{a priori} ambient space, nor node geometric information such as `coordinates'. Network embedding (also known as representation learning) is the process of assigning such an ambient space (called the \emph{latent} or \emph{embedding space}) to a network. This is typically done by mapping the nodes to a geometric space, such as a Euclidean space $\mathbb{R}^n$, while preserving some properties of the nodes, links, and/or network \cite{Cui2019}.  Overall, network embedding methods are used for learning a low-dimensional vector representation from a high-dimensional (as measured by the number of nodes) network. The relationships between nodes in the network are represented by their distance in the low-dimensional embedding space. Then, the low-dimension vector representation can be used for visualisation, and in a wide variety of downstream analyses, from network inference or link prediction to node classification or community detection. Moreover, network embedding can provide insights into the geometry of the underlying data. These insights can be useful for performance improvement, by working on a lower dimensional space, or exploiting the richer geometry of the embedding space. Finally, some downstream analyses, such as machine learning techniques, require a vector representation of the network. In this context, embedding network into a vector space is a prerequisite \cite{Nelson2019}.

Network embedding raises many challenges. First, a fundamental question is which network properties should be preserved by the embedding. For instance, the embedding space may preserve the intra-community similarity, the structural role similarity, or the similarity between nodes labels. A second challenge is related to the choice of dimension. The dimension of the embedding space will be a trade-off between two competing requirements: preserving the information encoded in the original network (favours high-dimensional space representations) and reducing the complexity or noise in the original network (favours low-dimensional space representations). Third, the scalability of network embedding methods is important: embedding methods applied to real networks face low parallelizability and data sparsity issues. Methods need to be efficient for networks of the size of typical modern-day network data sets, that is, up to several million nodes and edges \cite{NetworkRepositoryKONECT, NetworkRepositoryNetzschleuder, NetworkRepositoryUCI}. Lastly, the interpretation of the results of network embedding can be difficult \cite{Chari2021}.\\ 

For decades, dimensionality reduction methods based on factorisation matrices appeared as a relevant way to encode topological network information \cite{Jolliffe2016, Robinson1995, Ye2005}. These methods provided an initial set of network embedding techniques due to the success and ubiquity of network models. Over the past few years, there has been a significant surge in the number of embedding methods, making it challenging to navigate this fast-evolving field. The purpose of this review is to provide an overview based on a novel taxonomy that extends previous ones \cite{Hamilton2018, Chami2022} and describe current trends in network embedding.\\

First, we introduce the basic concept of network embedding and the state-of-the-art taxonomies of these methods. Next, we present our own taxonomy based on the common mathematical processes that underlie the embedding methods. This taxonomy aims to assist readers in navigating the field. We describe the two well-established classes of methods: the shallow embedding methods, and the deep learning methods. In addition to these two classical approaches, we include two sections dedicated to the higher-order network embedding methods and the emerging network embedding methods. These sections highlight current trends in the field, although the taxonomy is broad enough to integrate these new methods. Finally, we illustrate the wide range of network embedding applications, with one section devoted to the classical applications that include a user guideline and another section dedicated to the emerging applications that are currently growing in popularity.

\hfill
\section{Definitions and preliminaries}
\hfill

A \emph{network}, defined formally as a pair $G = (V, E)$, consists of a non-empty set $V$ of \emph{vertices} (or \emph{nodes}), and a set of \emph{edges} (or \emph{links}) $E$ connecting certain pairs of nodes. In the case of undirected networks, we can define $E$ as a subset of $\left\{\{u,v\} \mid u, v \in V\right\}$, and call $\{u,v\} \in E$ an \emph{undirected edge between vertices $u$ and $v$}, so that  $\{u,v\}=\{v,u\}$. In the case of directed networks, we can define $E \subseteq V \times V$, and call $(u,v) \in E$ a \emph{directed edge from vertex $u$ to vertex $v$}, so that $(u,v)\neq (v,u)$. If we agree on a labeling of the vertices, $V=\{v_1,\ldots,v_n, \ldots\}$, we can write $e_{ij} \in E$ for a vertex between $v_i$ and $v_j$ (undirected case) or from $v_i$ to $v_j$ (directed case). Depending on the network model, we can also add node or edge weights and types (see below). 

In its simplest form, a network embedding maps each node of a network to a space $X$, typically a Euclidean vector space $X=\mathbb{R}^d$ with $d \ll n$ the number of nodes. This space is called the \emph{latent space} or \emph{embedding space}. In the latent space, certain properties (of the nodes, edges, or the whole network) are preserved. Hence, a network embedding (into $X=\mathbb{R}^d$) is a mapping function 
\begin{align*}
  f \colon V & \rightarrow \mathbb{R}^{d} \\
  v_{i} & \mymapsto z_{i}.
\end{align*}
The embedding vector $z_i$ is expected to capture the topological properties of the original network while reducing the network dimension $n$. Network embedding methods can embed different components of the network. The previous definition is describing the most common embedding, namely node embedding method. In node embedding methods, each node of the network is embedded into an embedding space $X$, typically a reduced vectorial representation, that is, a mapping function $V \to X$. However, some methods handle edge embeddings $E \to X$, where each edge of the network is embedded into an embedding space $X$. Other embedding methods target subgraph or whole-network embedding, where the whole network, or some of its parts, are projected into an embedding space, such as a vector space. \\

To design efficient network embedding methods, several criteria need to be considered:
\begin{itemize}
    \item \textbf{Adaptability}: Embedding methods need to be applicable to different data and task, without, for instance, repeating a learning step.
    \item \textbf{Scalability}: Embedding methods need to process large-scale networks in a reasonable time.
    \item \textbf{Topology awareness}: The distance between nodes in latent space should reflect the connectivity and/or homophily (similar nodes in a network will be close in the embedding space) of the nodes in the original network. The homophily is the tendency of nodes to be connected to similar nodes.  
    \item \textbf{Low dimensionality}: Embedding methods should reduce the dimension of the network, by mapping a network with $n$ nodes to a $d$-dimensional space, with $d\ll n$.
    \item \textbf{Continuity}: The latent space should be continuous, which is beneficial in some tasks like classification \cite{Chen2018}.
\end{itemize}

As mentioned, while reducing the dimension, the embedding space should preserve some node, edge, and/or network properties. Focusing on node properties, the most common properties preserved by network embedding methods include:

\begin{itemize}
    \item \textbf{The first-order similarity} between two vertices, which is the pairwise similarity between the vertices. In other words, the weight of the edge between vertices defines a first-order similarity measure. Let $s_{v_{i}}$ (respectively $s_{v_{j}}$) be the first-order vector similarity associated with the node $v_{i}$ (resp.~$v_{j}$) to every other node in the network.
    \item \textbf{The second-order similarity} between two vertices, which considers the similarity of vertices in terms of neighbourhood structures. The second-order similarity between the nodes $v_{i}$ and $v_{j}$ is defined as the similarity between the first-order vectors $s_{v_{i}}$ and $s_{v_{j}}$. Higher-order similarities are based on the same idea. These second or higher-order similarities define structural equivalence between nodes.
    \item \textbf{The regular equivalence similarity}, which defines the similarity between vertices that share common roles in their neighbourhood, i.e., that have similar local network structures. For instance, if a node is a bridge between two communities, or if a node belongs to a clique. The regular equivalence aims to unveil the similarity between distant vertices which share common roles, in contrast to to common neighborhood.
    \item \textbf{The intra-community similarity}, which defines the similarity between vertices in the same community. The intra-community similarity aims to preserve the cluster structure information of the network.
\end{itemize}

Embedding methods are often designed to use specific types of networks as input. These network types include:
\begin{itemize}
    \item \textbf{Homogeneous networks}, which correspond to the standard definition of networks mentioned above $G=(V,E)$, where $V$ is a non-empty set of vertices (nodes) and $E$ a set of (directed, or undirected) edges (links). A more general setup, which allows multi-edges, is $G=(V,E,s,t)$ where $V \neq \emptyset$ and $E$ are arbitrary (vertex, edge) sets, and $s, t \colon E \to V$ are the source, respectively target, functions. Homogeneous networks can also be weighted: an edge, respectively a vertex, a weight function is a function $w_E \colon E \to X_E$, respectively $w_V \colon V \to X_V$, where $X_E$ and $X_V$ are weight sets, typically numeric $X_V=X_E=\mathbb{R}$. 
    \item \textbf{Heterogeneous networks.} In homogeneous networks, the nodes and the edges are all the same type. In a heterogeneous network, nodes and edges can have types. Formally, a heterogeneous network is a network $G = (V, E)$, associated with two type functions $\phi: V \rightarrow A$ and $\psi: E \rightarrow R$. These functions associate each node (respectively edge) to its type. More precisely, we define $A = \{a_{1}, a_{2}, ..., a_{\alpha}\}$, with $\alpha$ the number of node types, and $R = \{r_{1}, r_{2}, ..., r_{\beta}\}$, with $\beta$ the number of edge types. If $\lvert A \rvert =\lvert R \rvert = 1$, 
    the network is homogeneous.
    \item \textbf{Signed networks.} This is a particular case of a weighted homogeneous network with weights $\pm1$. Namely, $G = (V,E)$ is a network, and $\tau: E \rightarrow \{-1,1\}$ is a mapping function that associates a sign to each edge. 
    \item \textbf{Multilayer networks.} A multilayer network is a type of heterogeneous network where the nodes are grouped into layers, and the edges can connect nodes in the same, or different, layers. Formally, a multilayer network is a triplet $\mathcal{M} = (Y, G, \mathcal{G})$, where $Y$ is the layer index set, $G =\{ G_\alpha \mid \alpha \in Y\}$ are (homogeneous) networks $G_\alpha = (V_\alpha,E_\alpha)$, and $\mathcal{G} = \{\mathcal{G}_{\alpha\beta} \mid \alpha, \beta \in Y\}$ are bipartite networks $\mathcal{G}_{\alpha\beta}=(V_\alpha,V_\beta,E_{\alpha\beta})$ encoding the inter-layer connectivity. There is a rich literature on multilayer networks, with different special cases such as multiplex or temporal networks \cite{Bianconi2018, Boccaletti2014, DeDomenico2016, Kivelae2014, DeDomenico2013, DeDomenico2014}. The interested reader can refer to \cite{Bianconi2018} for an extended overview.
    \item \textbf{Temporal networks.} Temporal networks are specific cases of multilayer networks where the layers are ordered by time, that is, they represent the evolution of a graph over time \cite{Bianconi2018, masuda, Holme2012}.
    \item \textbf{Knowledge networks.} Knowledge graphs are defined as a set of triples $(u,r,v) \in V\times R\times V$, where the nodes $u$ and $v$ belong to the nodes set $V$, and they are connected by edges of type $r \in R$.
\end{itemize}

\hfill
\section{Existing Taxonomies of network embedding methods}
\hfill

The huge amount and variety of embedding methods \cite{Li2020} make their classification into a unified taxonomy a difficult task. The methods can indeed be sorted according to several criteria. We will briefly present some of the most common taxonomies. \\

A first way to classify network embedding methods is based on the type of networks used as input. Some authors thereby distinguish the methods designed for homogeneous or heterogeneous networks \cite{Li2020} . The same strategy can be used to classify embedding methods designed for static and temporal networks, or single and multilayer networks. Based on this taxonomy, it is possible to add a layer of complexity by considering the type of component that the methods embed in the vectorial space, i.e., the nodes, the edges, the subgraphs, or the whole network. 
Other authors use some properties of the network embedding process to classify the different methods \cite{Zhang2020}. For instance, the network embedding methods may be classified depending on the network properties they intend to preserve, in addition to the part of the network the methods is focused (on the nodes, edges, or the whole network). Focusing on nodes, three different types of property preservations can be defined, at different scales: microscopic, mesoscopic, and macroscopic properties. Methods preserving microscopic properties retain structural equivalences between nodes, such as first-order, second-order, or high-order similarities. They hence seek to preserve the homophily existing in the original network. Methods preserving mesoscopic properties focus on the regular equivalence between nodes, on intra-community similarity, or, more generally, on properties that are in between the close node neighbourhood and the whole network. Finally, methods preserving macroscopic properties  tend to preserve whole network properties, like the scale-freeness \cite{Feng2018}. Based on the same idea, a different taxonomy has been adopted by Cui et al.~\cite{Cui2019}. In this work, the authors discriminate the network embedding based on the information they wish to encode. The first class of methods, called structure and property preserving methods, preserves structural information like the neighbourhood or the community structures. The second class, called information-preserving methods, constructs the embedding using complementary information like node labels and types (for heterogeneous networks) or edge attributes. The third class, called advanced information-reserving method, gathers supervised methods that propose an end-to-end solution (learning process where all parameters are trained jointly) and use various complementary information to learn the embedding space. \\

In conclusion, multiple taxonomies have been proposed, based on several criteria: the property preserved by the properties preserved by the network embedding methods \cite{Zhang2020, Cui2019}, the type and the properties of input networks \cite{Chen2018, Li2020}, or based on mathematical considerations \cite{Goyal2018, Chen2020, Li2022, Chami2022}. Our approach for the review is based on mathematical considerations and similar to those in Chami et al.~\cite{Chami2022}. However, while Chami et al.~extended the encoder-decoder framework of Hamilton et al.~\cite{Hamilton2018} (see section 3) to include deep-learning methods as a special case. Herein, we have defined a more flexible approach to organize these methods that are not constraints by the encoder-decoder framework, while this framework can be a useful tool for understanding the methods, we believe that a more flexible approach will offer easier integration of new methods into this taxonomy. Significantly, our review includes higher-order network embedding methods that were not covered in \cite{Chami2022}. Note that our review presents a wide range of methods akin to a diverse set of `species' coexisting within a `zoo', hence the chosen title for our review. In this way, the objective of this new taxonomy is to be fine-grained, to offer a consensual view, and to be easily extended to integrate novel methods. In addition, our approach is independent of the scientific domain of development and application of the embedding methods.

\hfill
\section{Taxonomy of network embedding methods}
\hfill

Recently, important efforts have been made to produce general frameworks defining different embedding methods under a common mathematical formulation \cite{Hamilton2018, Yang2020, Chami2022}. Notably, Hamilton et al.~\cite{Hamilton2018} proposed an encoder-decoder framework to define embedding methods, following four components:
\begin{enumerate}
    \item A pairwise similarity function: $s_{\mathcal{G}} : V \times V \rightarrow \mathbb{R}^{+}$. \\
    This function defines the similarity measure between the nodes in the original (i.e., direct) network space.
    \item An encoder function: $\text{Enc} : V \rightarrow \mathbb{R}^{d}$. \\
    This function encodes the nodes into the embedding space. For instance, the node $v_{i} \in V$ is embedded into the vector $z_{i} \in \mathbb{R}^{d}$.
    \item A decoder function: $\text{Dec} : \mathbb{R}^{d} \times \mathbb{R}^{d} \rightarrow \mathbb{R}^{+}$. \\
    This function associates a similarity measure in the embedding space to each pair of embedding vectors.
    \item A loss function: $l : \mathbb{R} \times \mathbb{R} \rightarrow \mathbb{R}$. \\
    This function measures the quality of the pairwise reconstruction. The objective is to minimise the errors of the reconstruction as follows: $\text{Dec}(\text{Enc}(v_{i}), \text{Enc}(v_{j})) = \text{Dec}(z_{i}, z_{j}) \approx s_{G}(v_{i}, v_{j})$. Most approaches minimise an empirical loss function around a set of training nodes (noted $\mathcal{D}$) rather than the theoretical loss function.
    \begin{align}
         \mathcal{L} = \sum _{(v_{i},v_{j}) \in \mathcal{D}} l(\text{Dec}(z_{i}, z_{j}), s_{\mathcal{G}}(v_{i}, v_{j})) \, ,
    \end{align}
\end{enumerate}
Many embedding techniques align with this taxonomy, but a few of them are inaccurately described. For instance, higher-order network embedding techniques \cite{Zhou2006, Gui2016, Feng2019} do not only use pairwise similarity functions (see section 3.C). While this framework has been extended and enhanced to incorporate deep learning techniques \cite{Chami2022}, it still does not cover higher-order network embedding methods.

Another general framework has been proposed by Yang et al.~\cite{Yang2020} to classify heterogeneous network embedding (HNE) methods. The idea of this framework is to convert the homophily principle (similar nodes in a network will be close in the embedding space) into a generic objective function:
    \begin{align}
         \mathcal{J} = \sum_{v_{i}, v_{j} \in V} w_{v_{i} v_{j}} d(z_{i}, z_{j}) + \mathcal{J}_{R} \, .
    \end{align}
The term $w_{v_{i} v_{j}}$ denotes the proximity weight, $d(z_{i}, z_{j})$ is the embedding distance between the embedding vectors associated with the nodes $v_{i}$ and $v_{j}$, and $\mathcal{J}_{R}$ represents some additional objectives such as regularisers.\\

The taxonomy proposed in this work is based on a mathematical point of view, illustrated in Fig.~1, which splits the methods into two main classes depending on their depth: the shallow embedding methods (3.A), and the deep learning methods (3.B). We complement these two classes by including higher-order methods (3.C), which can be classified as either shallow embedding or deep learning methods, enabling us to spotlight these new types of methods.
In the next sections, we adopt the notation defined in section 1 for both the network and the associated embedding. In the following, we write $\lVert . \rVert _{F}$ for the Frobenius norm, and $\lVert . \rVert _{2}$ for the Euclidean norm.

\begin{figure*}[ht]
	\captionsetup{justification=centering}
    \centering
	\includegraphics[width=16cm,height=16cm,keepaspectratio]{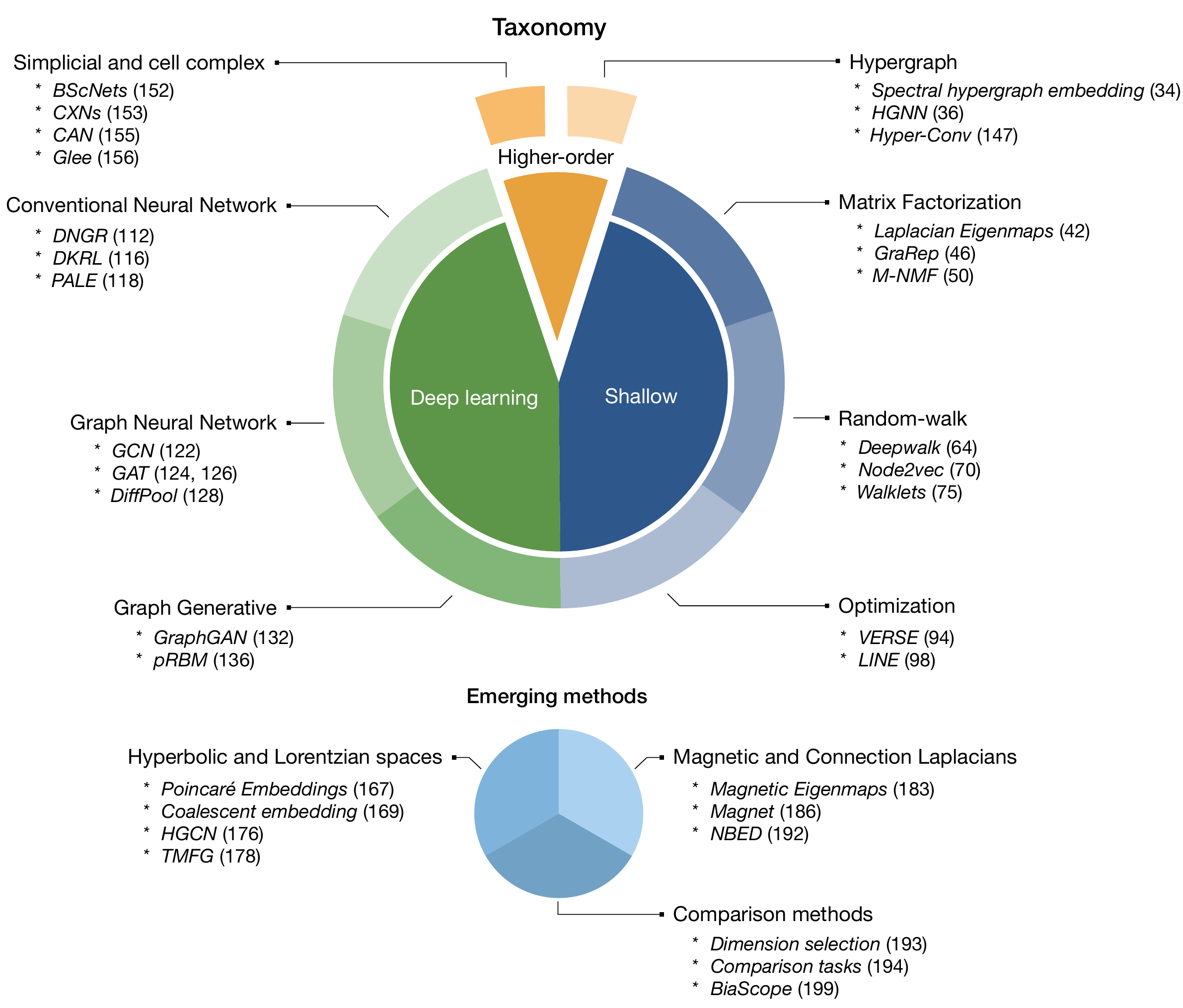}
    \caption{Pie charts describing the new taxonomy defined in this manuscript. In the top pie chart, the methods are divided into two main categories: shallow embedding methods and the deep learning methods, complemented by higher-order methods that can be either a shallow embedding or a deep learning methods. The bottom pie chart highlights the three major emerging groups of methods. Notably, these emerging groups of methods can be classified into our defined taxonomy due to its flexibility.}
\end{figure*}

\subsection{Shallow network embedding methods} 
\hfill\\

In this section we will consider the shallow network embedding methods, which are a set of methods with an encoder function that can be written as follows:
\begin{align}
    \text{Enc}(v_{i}) = {\bf Z v_{i}} \, ,
\end{align}
where ${\bf Z}$ corresponds to the matrix with the embedding vectors of all nodes, and ${\bf v_{i}}$ corresponds to the indicator vector associated with each node $v_{i}$ (vector of zeros except in position $i$, where the element is equal to $1$). In this case, the objective of the embedding process is to optimise the embedding matrix ${\bf Z}$ in order to have the best mapping between the nodes and the embedding vectors (Fig.~2). We define three major classes of shallow embedding methods based on different mathematical processes: the matrix factorisation methods, the random walk methods, and the optimisation methods.

\begin{figure*}[ht]
	\captionsetup{justification=centering}
    \centering
	\includegraphics[width=14cm,height=14cm,keepaspectratio]{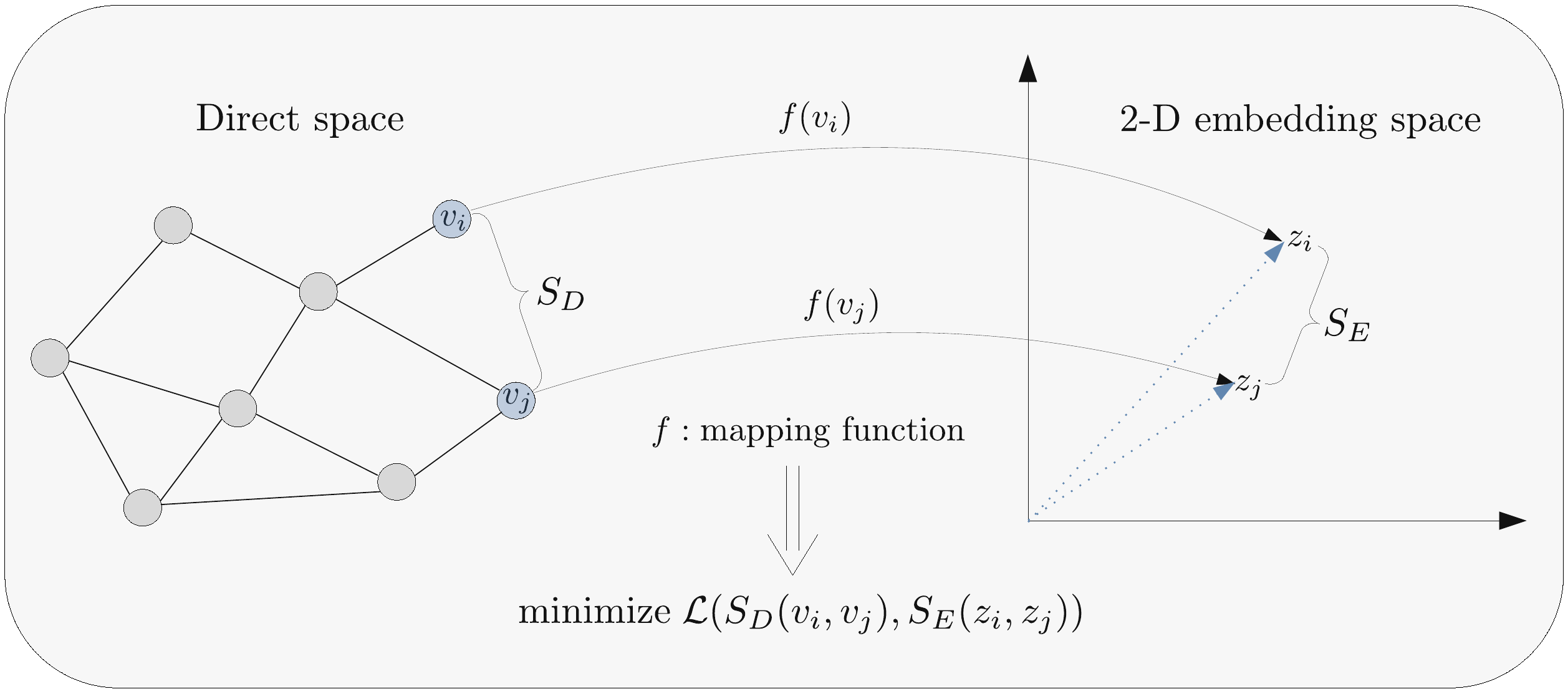}
    \caption{Shallow network embedding: To perform shallow network embedding, a network is projected into a low-dimensional vector space, such as a 2-dimensional embedding space. This projection is achieved using a mapping function $f$ that enables the mapping from the direct space to the embedding space. The mapping function $f$ is derived by optimizing a loss function $\mathcal{L}$, which aims to minimize the difference between the similarity measures of nodes in the direct space ($S_{D}$) and their equivalents in the embedded space ($S_{E}$) obtained through the decoder function.}
\end{figure*}

\subsubsection{Matrix factorisation methods}
\hfill\\

Matrix factorisation is based on the fact that a matrix, such as the adjacency matrix or the Laplacian matrix, can fully represent a network. 
This fact implies that existing methods from matrix algebra, such as matrix factorisation, can be used for network embedding. Network embedding methods based on matrix factorisation are directly inspired by linear dimensionality reduction methods, such as PCA \cite{Wold1987}, LDA \cite{Ye2005}, or MDS \cite{Chen2018}. Other methods are inspired by non-linear dimensionality reduction methods such as Isomap \cite{Samko2006}, which is an extension of MDS \cite{Chen2018}, LLE \cite{Roweis2000}, t-SNE \cite{Maaten2008}, or more recently UMAP \cite{McInnes2018}.
The factorisation process depends on the properties of the matrices. For positive semi-definite matrices, like graph Laplacians, the embedding can be obtained by eigenvalue decomposition. However, for unstructured matrices, like covariance matrices, 
gradient descent or Singular Value Decomposition (SVD) should be used to obtain the network embedding. Thereafter, we will describe the most common network embedding methods based on matrix factorisation.

\begin{itemize}
    \item {\bf Laplacian Eigenmaps (LE)} \cite{Belkin2003} aims to embed the network in such a way that two nodes close in the original network are also close in the low-dimensional embedding space, by preserving a similarity measure defined by the weight between nodes. In that way, we define the weight matrix denoted by $W$, where $W_{ij}$ encodes the weight between the nodes $i$ and $j$. The learning process is done by optimizing the following objective function:
        \begin{align}
             \mathcal{L} = \sum_{v_{i},v_{j} \in V} \text{Dec}(z_{i}, z_{j}) \cdot s_{\mathcal{G}}(v_{i}, v_{j}) \, ,
        \end{align}
    with $\text{Dec}(z_{i}, z_{j}) = \lVert z_{i} - z_{j} \rVert ^{2}_{2}$, and $s_{\mathcal{G}}(v_{i}, v_{j}) = W_{ij}$.\\
    We can introduce the Laplacian matrix $L$, defined as $L = D - W$, with $D_{ii} = \sum_{j}W_{ji}$. 
    The equation [4] can be written as:
        \begin{align}
             \mathcal{L} &= \frac{1}{2} \sum_{i,j} \lVert z_{i} - z_{j} \rVert ^{2}_{2} W_{ij} \notag \\ 
            \mathcal{L} &= \sum_{i,i} \lVert z_{i} \rVert ^{2}_{2} D_{ii} + \sum_{i,j} \lVert z_{i}z_{j} \rVert_{2} W_{ij} \notag \\
            \mathcal{L} &= \sum_{i,i} \lVert z_{i}z_{j} \rVert _{2}L_{ij} \notag \\
            \mathcal{L} &= \Tr(ZZ^{T}L) = \Tr(Z^{T}LZ) \, ,
        \end{align}
    with $Z = (z_{1}, z_{2}, ..., z_{n}) \in \mathbb{R}^{d \times n}$. The loss function needs to respect the constraint $Z^{T}DZ = I$ to avoid trivial solutions. The solution can be obtained by finding the matrix composed of the eigenvectors associated with the $d$ smallest eigenvalues of the generalized eigenvalue problem $LZ = \Lambda DZ$, with $\Lambda = diag([\lambda_{1}, \lambda_{2}, ..., \lambda_{n}])$ \cite{Ghojogh2019}. \\
    It is important to note that Laplacian Eigenmaps use a quadratic decoder function. This function does not preserve the local topology because the quadratic penalty penalises the small distance between embedded nodes.
  
    \item {\bf Cauchy Graph Embedding} \cite{Luo2011} aims to improve the previous method (Laplacian Eigenmaps), which does not preserve the local topology. Cauchy Graph Embedding use a different decoder function $\text{Dec}(z_{i}, z_{j}) = \frac{\lVert z_{i} -z_{j}\rVert ^{2}_{2}}{\lVert z_{i} -z_{j}\rVert ^{2}_{2} + \sigma^{2}} = 1 - \frac{\sigma^{2}}{\lVert z_{i} -z_{j}\rVert ^{2}_{2} + \sigma^{2}}$, with $\sigma^{2}$ representing the variance. Consequently, the loss function can be written as follows: 
        \begin{align}
             \mathcal{L} = \sum_{i,j} \frac{1}{\lVert z_{i} -z_{j}\rVert ^{2}_{2} + \sigma^{2}} W_{ij} \, ,
        \end{align}
    with the following constraints: $\sum_{i} z_{i} = 0$, and $Z^{T}Z = I$, where $Z = (z_{1}, z_{2}, ..., z_{n}) \in \mathbb{R}^{d \times n}$. The solution is obtained by an algorithm that mixes gradient descent and SVD.
    
    \item {\bf Graph factorisation} \cite{Ahmed2013} proposes a factorisation method that is designed for network partitioning. It learns an embedding representation that minimises the number of neighbouring vertices across the partition. The loss function can be written as follows:
        \begin{align}
            \mathcal{L} = \sum_{v_{i},v_{j} \in V} \lVert \text{Dec}(z_{i}, z_{j}) - s_{\mathcal{G}}(v_{i}, v_{j}) \rVert ^{2}_{2} + \frac{\lambda}{2}\sum_{v_{i} \in V} \lVert  z_{i} \rVert ^{2}_{2} \, ,
        \end{align}
    with $\text{Dec}(z_{i}, z_{j}) = z_{i}^{T}z_{j}$, $s_{\mathcal{G}}(v_{i}, v_{j}) = W_{ij}$, and $\lambda$ a regularisation parameter. Notably, this method is scalable and can deal with networks with millions of vertices and billions of edges.
        
    \item {\bf GraRep} \cite{Cao2015} extends the skip-gram model \cite{Mikolov2013} to capture higher-order similarity, i.e., $k$-step neighbours of nodes. The value of $k$ is chosen such as $1\leq k \leq K$, with $K$ the highest order. GraRep is also motivated by the Noise-Contrastive Estimation (NCE) approximation \cite{Gutmann2010} which consists in learning a model that converges to the objective function. INCE trains a binary classifier to distinguish between node samples coming from the similarity distribution $s_{\mathcal{G}}$ and node samples generated by a noise distribution over the nodes. Grarep defines its $k$-step loss function as follows:
        \begin{align}
            \mathcal{L}_{k} =& \sum_{v_{i} \in V} \Big(\sum_{v_{j} \in V} T^{k}_{i,j}\log(\sigma(x_{i}^{T}x_{j})) \notag \\
            &+ \lambda \mathbb{E}_{v_{j} \sim p_{k}(V)}[\log(\sigma(-x_{i}^{T}x_{j}))]\Big) \, ,
        \end{align} 
    where the matrix $T$ represents the transition matrix, defined as $T = D^{-1}A$, with $A$ the adjacency matrix, and $D$ the degree matrix. The vectors $x_{i}$ and $x_{j}$ are the vector representations of the nodes $v_{i}$ and $v_{j}$ in the direct space. The term $\mathbb{E}_{v_{j} \sim p_{k}(V)}$ is the expectation of the node $v_{j}$, obtained by negative sampling. The expectation follows the noise distribution over the nodes in the network, denoted by $p_{k}(V)$. The parameter $\lambda$ indicates the number of negative samples, and $\sigma(.)$ is the sigmoid function defined as $\sigma(x) = (1 + e^{-x})^{-1}$. GraRep reformulates its loss function minimisation into a matrix factorisation problem. Each $k$-step term is computed from the matrix $X^{k}$ defined as $X^{k}_{ij} = \max([\log\big(\frac{T^{k}_{ij}}{\sum_{m} T^{k}_{mj}}\big) - \log(\beta)], 0)$. Then, the low-dimensional representation of the matrix $C^{k}$ is constructed from the Singular Value Decomposition: $\text{SVD}(X^{k})$. Finally, the final representation is obtained by concatenating all order-term matrices, $C = [C^{1}, C^{2}, ..., C^{K}]$.        
        
    \item {\bf High-Order Proximity preserved Embedding (HOPE)} \cite{Ou2016} has been developed to encode higher-order similarity of large-scale networks while also capturing the asymmetric transitivity, i.e., going from node $v_{i}$ to node $v_{j}$ can be different from going from node $v_{j}$ to node $v_{i}$. HOPE can hence deal with directed networks. The loss function is equal to:
        \begin{align}
            \mathcal{L} = \sum_{v_{i},v_{j} \in V} \lVert \text{Dec}(z_{i}, z_{j}) - s_{\mathcal{G}}(v_{i}, v_{j}) \rVert ^{2}_{2} \, ,
        \end{align}
    with $\text{Dec}(z_{i},z_{j}) = z_{i}^{T}z_{j}$ and $s_{\mathcal{G}}(v_{i}, v_{j})$ denoting any similarity measure between $v_{i}$ and $v_{j}$. The authors of HOPE introduce a general factorisation in which the similarity measure can be factorised in one matrix associated with the global similarity $M_{g}$ and another matrix associated with the local similarity $M_{l}$. So, the similarity matrix can be expressed as $S = M_{g}^{-1}M_{l}$, where both local and global similarities are polynomial sparse matrices. This also enables HOPE to use efficient SVD decomposition for embedding large-scale networks. The authors considered different similarity measures such as Katz index ($S^{katz} = (I-\beta A)^{-1} (\beta A)$), Rooted PageRank ($S^{RPR} = (I-\alpha T)^{-1} ((1-\alpha)I)$), common neighbours ($S^{CN} = I(A^{2})$), or Adamic-Adar ($S^{AA} = I(ADA)$), where $A$ indicates the adjacency matrix, $T$ represents the transition matrix, $\alpha$ a value $\in [0,1)$, and $\beta$ a value less than the spectral radius of the adjacency matrix.\\
    
    \item {\bf Modularised Nonnegative Matrix factorisation (M-NMF)} \cite{Wang2017a} aims to obtain an embedding representation aware of the community structure of the original network while maintaining the microscopic information from the first-order and second-order similarities. Let us define the similarity measure $S = S^{(1)} + \eta S^{(2)} \in \mathbb{R}^{n\times n}$, where $S^{(1)}$ is the first-order similarity matrix, for instance, $S^{(1)}_{ij} = A_{ij}$ with $A$ the adjacency matrix, and $S^{(2)}$ is the second-order similarity matrix, defined as $S^{(2)}_{ij} = \frac{\mathcal{N}_{i}\mathcal{N}_{j}}{\lVert \mathcal{N}_{i} \rVert _{2} \lVert \mathcal{N}_{j} \rVert _{2}}$, with $\mathcal{N}_{i} = (S^{(1)}_{i1}, S^{(1)}_{i2}, \ldots, S^{(1)}_{in})$ the first-order similarity vector of the node $i$. The parameter $\eta$ is the weight of the second-order term (often chosen equal to 5 \cite{Wang2017a}). The embedding of the microscopic structure can be expressed in the NMF framework as the following optimisation problem:
        \begin{align}
            \min_{M, U} \;\; \lVert S - MU^{T} \rVert ^{2}_{F} \; ; \;\;\; M > 0 \;,\; U > 0 ,
        \end{align}
    with $M \in \mathbb{R}^{n\times d}$ and $U \in \mathbb{R}^{n\times d}$ two non-negative matrices; $U_{i}$ is the embedding of the node $i$.\\
    The community structure is obtained with modularity maximisation, which is expressed for two communities as $Q = \frac{1}{4m} \sum_{ij} (A_{ij} - \frac{k_{i}k_{j}}{2m})h_{i}h_{j}$, with $k_{i}$ the degree of the node $i$, $h_{i}$ is equal to $1$ if the node $i$ belongs to the first community, otherwise is equal to $-1$, and $m$ is the total number of edges. Let us define $B$ such as $B_{ij} = A_{ij} - \frac{k_{i}k_{j}}{2m}$, so the modularity becomes $Q = \frac{1}{4m} h^{T}Bh$, where $h\in \mathbb{R}^{n}$. The generalisation of the modularity optimisation problem for $k$ communities is defined as:
        \begin{align}
            \min_{H} \;\; -\beta \Tr{(H^{T}BH)} \; ; \;\;\; \Tr{(H^{T}H)} = n ,
        \end{align}
    with $H \in \mathbb{R}^{n\times k}$, $\beta$ is a positive parameter.
    The second equation imposes the association of each node to one community. The two models are combined using a term that uses the community structure to guide the node representation learning process. Formally, we define $C \in \mathbb{R}^{k\times d}$ as the community representation matrix; $C_{r}$ as the representation of the community $r$, and $U_{i}C_{r}$ represents the propensity of the node $i$ to belong to the community $r$. So the last term to optimise is equal to $\alpha \lVert H - UC^{T} \rVert ^{2}_{F}$, with the constraint that $C > 0$, and $\alpha$ a positive parameter. Finally, the equation to be optimised is the following one:
        \begin{gather}
            \min_{M, U, H, C} \;\; \lVert S - MU^{T} \rVert ^{2}_{F} -\beta \Tr{(H^{T}BH)} + \alpha \lVert H - UC^{T} \rVert ^{2}_{F} \notag \\
            M > 0 \;,\; U > 0 \;,\; C > 0 \;,\; \Tr{(H^{T}H)} = n \, .
        \end{gather}
    Due to the non-convex behavior of the previous function, a non-trivial optimisation process has been developed \cite{Wang2017a}.
    
    \item {\bf Text-Associated DeepWalk (TADW)} \cite{Yang2015} aims to integrate text data information into the network embedding process. The authors first prove that the learning process used in the Deepwalk embedding method (see the section about Random walk network embedding methods) is equivalent to the optimisation of a matrix factorisation problem, $M = W^{T}H$, with $M \in \mathbb{R}^{n\times n}$ the matrix of the original network, $W \in \mathbb{R}^{d\times n}$ the weight matrix, and $H \in \mathbb{R}^{d\times n}$ the factor matrix. The factorisation matrix problem is the following: 
        \begin{align}
            \min_{W, H} \;\; \lVert M - W^{T}H \rVert ^{2}_{F} + \frac{\lambda}{2}(\lVert W \rVert ^{2}_{F} + \lVert H \rVert ^{2}_{F}) \, .
        \end{align}
    The idea of TADW is to take into account a text factor matrix $T$ into the decomposition, such that $M = W^{T}HT$, with $M \in \mathbb{R}^{n\times n}$, $W \in \mathbb{R}^{d\times n}$, $H \in \mathbb{R}^{d\times k}$, and $T \in \mathbb{R}^{k\times n}$. The new factorisation matrix problem is:
        \begin{align}
            \min_{W, H} \;\; \lVert M - W^{T}HT \rVert ^{2}_{F} + \frac{\lambda}{2}\Big(\lVert W \rVert ^{2}_{F} + \lVert H \rVert ^{2}_{F}\Big) \, .
        \end{align}
    The optimisation process is obtained with the gradient descent algorithm introduced by H. Yu et al.~\cite{Yu2014}.

    \item {\bf Other matrix factorisation methods}: The methods detailed above are some of the most common ones, and there are used as basis for alternative or extended methods. Notably, several strategies propose variations of the Laplacian Eigenmaps. For instance, the Locality Preserving Properties method \cite{He2004} uses a linear approximation of LE. The method Structure-Preserving Embedding \cite{Shaw2009} extends LE by including connectivity structure similarity as a constraint during the learning process. Similarly, Augmented Relational embedding \cite{Lin2005} modifies the Laplacian matrix to integrate feature information. \\ 
    Spectral techniques such as Label informed attributed Network Embedding \cite{Huang2017} are also promising for preserving node structure similarities and the correlations between their labels. \\
    Some methods dedicated to multi-class node classification have also been developed. These methods can be seen as variations of the TADW method. For instance, the method Homophily, Structure, and Content Augmented \cite{Zhang2016} adds a regularisation term to the objective function of TADW to enforce the structure homophily existing between nodes in the network. Max-Margin DeepWalk \cite{Tu2016} adds a multi-class SVM to integrate labelling information of the nodes. Discriminative Matrix factorisation \cite{Zhang2016a} uses a linear classifier trained on labelled nodes to complement the TADW objective function. \\
    A large number of other embedding methods based on matrix factorisation have been developed. For instance, several embedding methods applied to knowledge graphs, use matrix factorisation (or tensor factorisation). These methods can be defined as relation learning methods. We can mention some of the most common ones, such as RESCAL \cite{Nickel2012}, DistMult \cite{Yang2015a} and ComplEx \cite{Trouillon2016}. DistMult is a special case of RESCAL developed to reduce overfitting and ComplEx extents DistMult to complex matrices. \\
    Finally, matrix factorisation methods can extract network embeddings from a time-dependent node similarity measure inspired by dynamical systems and control theory notions \cite{schaub2019multiscale}. 
\end{itemize}

\subsubsection{Random walk methods}
\hfill\\

The idea behind random walk embedding is to encode the scores of the random walk into an embedding space. Most methods use ideas initially developed in the Deepwalk paper \cite{Perozzi2014}. In this section, we describe the most common methods and some of their extensions.

\begin{itemize}
    \item {\bf Deepwalk} \cite{Perozzi2014} is a scalable network embedding method that uses local information obtained from truncated random walks to learn latent representations. Deepwalk treats the walks as the equivalent of sentences. The process is inspired by the famous word2vec method \cite{Mikolov2013, Mikolov2013a}, in which short sequences of words from a text corpus are embedded into a vectorial space. The first step of Deepwalk consists in generating sequences of nodes from truncated random walks on the network. Then, the update procedure consists in applying the skip-gram model \cite{Mikolov2013} on the sequences of nodes, in order to  maximise the probability of observing a node neighbour conditioned on the node embedding. The loss function is defined as follows:
        \begin{align}
            \min_{\phi} \;\; -\log\Big(\mathbb{P}(\{v_{i-w}, ..., v_{i+w}\} \; \setminus \; v_{i} \; | \; \phi(v_{i}))\Big) \, ,
        \end{align}
    with $w$ indicating the window size (in terms of node sequence), and $\phi: V \rightarrow \mathbb{R}^{d}$ indicating the mapping function. We can also see $\phi \in \mathbb{R}^{n \times d}$ as the matrix of the embedding representation of the nodes. The skip-gram model transform the equation [15] as follows:
        \begin{align}
            \min_{\phi} \;\; -\log\Big(\prod_{j = i-w}^{i+w} \mathbb{P}(v_{j} \; | \; \phi(v_{i})\Big) \, .
        \end{align}
    Then, the hierarchical softmax function \cite{Mnih2008} is applied to approximate the joint probability distribution as:
        \begin{align}
            \mathbb{P}(v_{j} \; | \; \phi(v_{i})) &= \prod_{l = 1}^{\log(n)}\mathbb{P}(b_{l} \; | \; \phi(v_{i})) \notag \\
           &= \prod_{l = 1}^{log(n)} \frac{1}{1 + \exp(-\phi(v_{i})\psi(b_{l}))} \, ,
        \end{align}
    where $v_{j}$ is defined by a sequence of tree nodes $(b_{0}, b_{1}, ..., b_{\log(n)})$, with $b_{0}$ the root of the tree, and $b_{\log(n)}$ the node $v_{j}$. Notably, similarly to the TADW method, Deepwalk is equivalent to the following matrix factorisation problem: $M = W^{T} H$ \cite{Yang2015, Qiu2018}, with $W \in \mathbb{R}^{d\times n}$ the weight matrix, and $H \in \mathbb{R}^{d\times n}$ the factor matrix. Several extensions of the deepwalk have been adapted for multilayer networks \cite{Cen2019a, Dursun2020}.

    \item {\bf node2vec} \cite{Grover2016} is a modified version of Deepwalk, with two main changes. First, node2vec uses a negative sampling instead of a hierarchical softmax for normalisation. This choice improves the running time. Second, node2vec uses a biased random walk that offers more flexible learning with control parameters. The biased random walk can be described as:  
        \begin{align}
            \mathbb{P}(c_{i} = x \; | \; c_{i-1} = y) = \left\{
                                                            \begin{array}{ll}
                                                            \frac{\pi_{yx}}{Z} & \mbox{if $(y, x)$ $\in$ $E$,} \\
                                                            \;\; 0 & \mbox{otherwise,}\\
                                                            \end{array}
                                                            \right.
        \end{align}
    where $\pi_{yx}$ is the unnormalised transition probability between node $y$ and node $x$, and $Z$ is the normalizing constant. The variable $\pi$ is defined as follows:
        \begin{align}
            \pi_{yx} = \left\{
                                \begin{array}{ll}
                                \frac{1}{p}\omega_{yx} & \mbox{if $d_{tx} = 0$} \\
                                \omega_{yx} & \mbox{if $d_{tx} = 1$} \\
                                \frac{1}{q}\omega_{yx} & \mbox{if $d_{tx} = 2$} \
                                \end{array}
                                \right. \, ,
        \end{align}
    where $\omega_{yx}$ is the weight of the edge between the node $y$ and the node $x$, and $d_{tx}$ is the shortest path between the node $x$ and the node $t$, which is the node reached before the node $y$. The parameters $p$ and $q$ are two control parameters of the random walk. The return parameter $p$ controls the likelihood of immediately revisiting a node in the walk, while the in-out parameter $q$ controls the likelihood of visiting a node in the neighbourhood of the node that was just visited. Both parameters control if the random walk follows a Breadth-First Sampling (BFS) strategy or a Depth-First Sampling (DFS) strategy. The first strategy preserves the structural equivalence of the nodes, the second one preserves their homophily. Recently, Multinode2vec \cite{Wilson2018} and PMNE \cite{Liu2017a}, two extensions of node2vec, adapted the random walk process to multilayer networks.
    
    \item {\bf HARP} \cite{Chen2018a} is an algorithm that was developed to improve the Deepwalk and node2vec embedding methods. The idea is to capture the global structure of an input network by recursively coalescing edges and nodes of the network into smaller networks with similar structures (see also section 4.B on network compression). The hierarchy of these small networks is an appropriate  initialisation for the network embedding process, because it directly express a reduced-dimension version of the input network while preserving its global structure. The final embedding is obtained by propagating the embedding of the smallest network through the hierarchy.
    
    \item {\bf Discriminative Deep Random Walk (DDRW)} \cite{Li2016} is particularly suitable  for the network classification task. It can be seen as a Deepwalk extension that considers the label information of nodes. 
    To do so, DDRW jointly optimises the Deepwalk embedding loss function and a classification loss function. The final loss function to optimise is defined as:
        \begin{gather}
            \mathcal{L} = \eta \mathcal{L_{DW}} + \mathcal{L_{C}} \, , \\
            \mathcal{L_{C}} = C \sum_{i = 1}^{n} (\sigma(1 - y_{i} \beta^{T} z_{i}))^{2} + \frac{1}{2} \beta^{T}\beta \, ,
        \end{gather}
    where $\eta$ is a weight parameter, and $\sigma$ the Heaviside function, i.e.~$\sigma(x)=x$ for $x > 0$ and  $\sigma(x)=0$ otherwise. Moreover, $z_{i}$ is the embedding vector of the node $v_{i}$, $y_{i}$ is the label of the node $v_{i}$, $C$ is the regulariser parameter, and $\beta$ the subsequent classifier.
    
    \item {\bf Walklets} \cite{Perozzi2017}. Given the observation that Deepwalk can be derived from a matrix factorisation containing the powers of the adjacency matrix \cite{Yang2015b}, it appears that Deepwalk is biased towards lower powers of the adjacency matrix, which correspond to short walks. This can become a limitation when higher-order powers are the most appropriate representations, for instance to embed the regular equivalence between nodes. To bypass this issue, Walklets propose to learn the embedding from a multi-scale representation. This multi-scale representation is sampled from successive higher powers of the adjacency matrix obtained from random walks. Then, after partitioning the representation by scale, Walklets learns the representation of each node generated for each scale.
    
    \item {\bf Struct2vec} \cite{Ribeiro2017} aims to capture the regular equivalence between nodes in a network. In other words, two nodes that have identical local network structures should have the same embedding representation. 
    The construction of the embedding representation is based on different steps. The first step is to determine the structural similarity between each pair of nodes for different neighbourhood sizes. The structural similarity between the nodes $v_{i}$ and $v_{j}$, when considering their $k$-hop neighbourhoods (all nodes at a distance less or equal to $k$ and all edges among them), is defined as follows:
        \begin{gather}
            d_{k}(v_{i}, v_{j}) = d_{k-1}(v_{i}, v_{j}) + g(s(R_{k}(v_{i})), s(R_{k}(v_{j}))) \; ; \; \notag \\
            k \geq 0 \;\; \text{and} \;\; |R_{k}(v_{i})| , |R_{k}(v_{j})| > 0 \, ,
        \end{gather}
    where $R_{k}(v_{i})$ is the set of nodes at a distance less or equal to $k$ from the node $v_{i}$, $s(S)$ is the ordered degree sequence of a set of nodes $S$, and $g(S_{1}, S_{2})$ is a distance measure between the two ordered degree sequences $S_{1}$ and $S_{2}$. The distance used is the Dynamic Time Warping \cite{Salvador2007} and by convention $d_{-1} = 0$. \\
    This procedure produces a hierarchy of structural similarities between nodes of the network. The hierarchy is used to create a weighted multi-layer network, in which layers represent node similarities for different levels of the hierarchy. The edge weights between node pairs are inversely proportional to their structural similarity. After that, a biased random walk process is applied to the multilayer network to generate sequences of nodes. The sequence of nodes are used to learn a latent representation with the skip-gram process.
    
    \item {\bf Other random walk methods}: Some methods are designed to integrate additional information in the learning process. For instance, SemiNE \cite{Li2017b} is a semi-supervised extension of Deepwalk that takes into account node labels. GENE \cite{Chen2016} also integrates node labels. Node labels, as well as additional node contents, are also integrated into TriDNR \cite{Pan2016}. \\
    SNS \cite{Lyu2017} is another method that aims to preserve structural similarity in the embedding representation. SNS measures the regular equivalence between nodes by representing them as a graphlet degree vectors: each element of the graphlet degree vector represents the number of times a given node is touched by the corresponding orbit of graphlets. \\
    Random walk approaches are widely used for Heterogeneous Network Embedding (HNE). Examples of HNE method include MRWNN \cite{Wu2016}, SHNE \cite{Zhang2019b}, HHNE \cite{Wang2019}, GHE \cite{Chen2017}, JUST \cite{Hussein2018}, HeteSpaceyWalk \cite{He2019}, and TapEm \cite{Park2019}. The interested reader can refer to Yang et al.~\cite{Yang2020} for a detailed review of HNE. Finally, random walks are also often used for metapath-based methods, another set of methods relevant for network embedding. This set of methods includes Metapath2vec \cite{Dong2017}, HIN2vec \cite{Fu2017}, HINE \cite{Huang2017a}, or, more recently, HERec \cite{Shi2019}.
\end{itemize}

\subsubsection{Optimisation methods}
\hfill\\

The previous two methods involve two distinct mathematical processes: matrix factorisations, and random walks. (Matrix factorisation is a common mathematical operation, while random walk encompasses several methods that share a common principle.) However, there are additional embedding techniques that do not fit into either category, but they do share a common objective of optimising a loss function. In essence, these methods use a broad range of mathematical processes but ultimately involve an optimisation step, which is usually achieved through gradient descent. As a result, these approaches can be viewed as hybrid methods that all utilise a shared optimisation step.\\
The most important step in optimisation methods is to define a loss function that encodes all the properties that should be preserved through the embedding. This loss function often gathers similarities between nodes in the direct space, together with some regulariser terms that depend on network features that we want to preserve. The embedding representation is obtained based on the optimisation of this loss function. We will present the most common optimisation-based network embedding methods and some of their extensions.

\begin{itemize}
    \item {\bf VERtex Similarity Embeddings (VERSE)} \cite{Tsitsulin2018} is a versatile network embedding method that accepts any network similarity measure. Let $G$ be a network with an associated similarity measure $s_{\mathcal{G}}~:~V \times V \rightarrow \mathbb{R}^{+}$. The VERSE method constructs the embedding representation of the network $G$, noted $Z \in \mathbb{R}^{d \times n}$, associated with a similarity measure in the embedding space $\text{Dec}~:~\mathbb{R}^{d} \times \mathbb{R}^{d} \rightarrow \mathbb{R}^{+}$. Each column of the matrix $Z$ is the embedding vector $z_{i}$ of the node $v_{i}$. The embedding representation is based on the optimisation of a loss function, noted $\mathcal{L}$, corresponding to the Kullback-Leibler divergence between the similarity matrix in the direct space (i.e.~original network) and the similarity matrix in the embedding space:
        \begin{align}
            \mathcal{L} &= \sum\limits_{i = 1}^{n}s_{\mathcal{G}}(v_{i}, .) \cdot \ln\left(\frac{s_{\mathcal{G}}(v_{i}, .)}{\text{Dec}(v_{i}, .)}\right) \notag \\
            &= -\sum\limits_{i = 1}^{n}s_{\mathcal{G}}(v_{i}, .) \cdot \ln({\text{Dec}(v_{i}, .)}) + C \, ,
        \end{align}
    with $s_{\mathcal{G}}(v_{i}, .)$ (resp. $\text{Dec}(v_{i}, .)$) the similarity vector between the node $v_{i}$ and all the other nodes of the network in the direct space (resp. embedding space). Notably, $\sum_{j = 1}^{n} s_{\mathcal{G}}(v_{i}, v_{j}) = \sum_{j = 1}^{n} \text{Dec}(v_{i}, v_{j}) = 1$. Moreover, $C = \sum_{i = 1}^{n}s_{\mathcal{G}}(v_{i}, .) \cdot \ln({s_{\mathcal{G}}(v_{i}, .)})$ defines a constant that does not affect the optimisation algorithm, and can therefore be neglected. The vector $s_{\mathcal{G}}(v_{i}, .)$ corresponds to the vector associated with the node $v_{i}$ in the similarity matrix defined in the direct space. As stated above, the similarity matrix in the direct space (network) can be defined by several measures. The authors proposed three different similarity matrices: the adjacency matrix, the SimRank similarity matrix \cite{Jeh2002}, and a similarity matrix based on Random Walk with Restart.
        The vector $\text{Dec}(v_{i}, .)$ corresponds to the vector associated with the node $v_{i}$ in the similarity matrix defined in the embedding space. The vector $\text{Dec}(v_{i}, .)$ can also be seen as the similarity vector between the vectors $z_{i}$ and $z_{j}$ with $j \neq i, j \in \llbracket 1,n \rrbracket$, where $n$ is the number of nodes in the network. The vectors gathered in the similarity matrix in the embedding space are defined by the following equation:
        \begin{align}
            \text{Dec}(v_{i}, .) = \frac{\exp(z_{i}^{T} Z)}{\sum\limits_{j = 1}^{n} \exp(z_{i}^{T}  z_{j})} \, .
        \end{align}
        The node embedding is obtained by optimizing the loss function with a gradient descent algorithm. Usually, the embedding vectors are initialized with a normal distribution with a mean equal to zero. Because the Kullback-Leibler optimisation is a time-consuming process, a negative sampling procedure, such as NCE (Noise Contrastive Estimation) \cite{Gutmann2010, Mnih2012} is often used. Recently, an extension of VERSE to heterogeneous multiplex networks, named MultiVERSE, has been developed \cite{PioLopez2020}. \\
        
    \item {\bf Large Scale Information Network Embedding (LINE)} \cite{Tang2015} aims to embed both first-order and second-order similarities. 
        \begin{itemize}
            \item The embedding space of the first-order similarity can be obtained by the following  optimisation algorithm. \\
            Let us define the  theoretical expected probability as:
                \begin{align}
                    p_{1}(v_{i},v_{j}) = \frac{1}{1 + \exp(-z_{i}^{T}z_{j})} \; ; \; z_{i} \in \mathbb{R}^{d} \, ,
                \end{align}
            and the  empirical probability  as:
                \begin{align}
                    p^{*}_{1}(v_{i},v_{j}) = \frac{w_{ij}}{W} \; ; \; W = \sum_{(i,j) \in E} w_{ij} \, .
                \end{align}
            The main goal of LINE is to minimise the error between the theoretical expected probability $p_{1}$ and the empirical probability $p^{*}_{1}$. To do so, the loss function $O_{1}$ minimises the distance $d(p^{*}_{1}(.,.),p_{1}(.,.))$ by using the Kullback-Leibler divergence. Hence, $O_{1}$ can be written as follows:
                \begin{align}
                    O_{1} = -\sum_{(i,j) \in E} w_{ij} \log(p_{1}(v_{i},v_{j})) \, .
                \end{align}
            \item The embedding space of the second-order similarity is obtained with the optimisation process described as follows. Let us define the  theoretical expected probability as:
                \begin{align}
                    p_{2}(v_{j} \; | \; v_{i}) = \frac{\exp(z_{j}^{T}z_{i})}{\sum_{k = 1}^{n} \exp(z_{k}^{T}z_{i})} \; ; \; z_{i} \in \mathbb{R}^{d} \, .
                \end{align}
            The empirical probability is defined as:
                \begin{align}
                    p^{*}_{2}(v_{j} \; | \; v_{i}) = \frac{w_{ij}}{d_{i}} \; ; \; d_{i} = \sum_{i \in N(i)} w_{ik} \, ,
                \end{align}
            where $N(i)$ is the neighborhood of the node $i$, and $d_{i}$ defines the out-degree of the node $i$. The idea is again to minimise the error between the theoretical expected probability and the empirical probability. To do so, the loss function $O_{2}$ minimises the distance $d(p^{*}_{2}(. \; | \; v_{i}), p_{2}(. \; | \; v_{i}))$ by using the Kullback-Leibler divergence. Hence, $O_{2}$ can be written as follows: 
                \begin{align}
                    O_{2} = -\sum_{(i,j) \in E} w_{ij} \log(p_{2}(v_{j} \; | \; v_{i})) \, .
                \end{align}
        \end{itemize}
        The first and second-order node representations are computed separately, and both first and second-order embedding representations are concatenated for each node. \\
        
    \item {\bf Transductive LINE (TLINE)} \cite{Zhang2016b} is a transductive version of LINE that also uses node labels to train a Support Vector Machine (SVM) classifier for node classifications. Both node embedding and the SVM classifier are optimised simultaneously, in order to make full use of the label information existing in the network. Notably, TLINE as LINE permits fast embedding of large-scale networks by using edge sampling and negative sampling in the stochastic gradient descent process. The method optimises a loss function $O_{T}$ composed of the same loss functions as LINE, $O_{1}$ and $O_{2}$ to embed both first and second-order similarity, and the SVM loss function $O_{SVM}$, that is, 
        \begin{gather}
            O_{T} = O + \beta O_{SVM} \, , \\
            O_{SVM} = \sum\limits_{i = 1}^{n} \sum\limits_{k = 1}^{K} \max(0, 1 - y_{i}^{k} w_{k}^{T} z_{i}) + \lambda \lVert w_{k} \rVert^{2} \, ,
        \end{gather}
        where $\beta$ is a trade-off parameter between LINE and SVM, $n$ the number of nodes, $K$ the number of label types in the network, $z_{i}$ the embedding vector representation of the node $v_{i}$, $w_{k}$ the parameter vector of the label class $k$, and $y_{i}^{k}$ is equal to 1 if the node $v_{i}$ is in the class $k$.
        
    \item {\bf Other optimisation methods}: A wide range of optimisation methods are applied to heterogeneous networks. Many of them are similar to LINE and optimise first and second-order similarities. In PTE (Predictive Text Embedding) \cite{Tang2015a}, the loss function is divided into several loss functions, each associated with one network of the heterogeneous network. The APP (Asymmetric Proximity Preserving) \cite{Zhou2017} network embedding method is similar to VERSE, and captures both asymmetric and high-order similarities between node pairs thanks to an optimisation process over random walk with restart results. \\
    Recently, many network embedding methods have been adapted or designed for multilayer networks, and a significant portion of them are based on an optimisation process \cite{Zitnik2017, Xu2017a, Zhang2018, Bagavathi2018, An2021}.
    In addition, several embedding methods applied to knowledge graphs are optimisation methods. These methods are often called relation learning methods. We can mention some of the most common ones, like the translation-based methods first defined by Bordes et al.~\cite{Bordes2013}. This method, named TransE, embeds multi-relational data that uses directed graphs. Edges can be defined by three elements: the head node ($h$), the tail node ($t$), and the edge label ($l$). The embedding vector of the tail node $t$ should be close to the embedding vector of the head node $h$, plus some vector that depends on the relationship $l$. This approach constructs the embedding representation by optimizing a loss function that integrates these three elements. This method has given rise to several alternative methods: TransH \cite{Wang2014} that improves TranSE for reflexive/one-to-many/many-to-one/many-to-many relationships, TransR \cite{Lin2015} that builds entity and relation embeddings in separate entity space and relation space (in contrast to the two previous methods), and TransD \cite{Ji2015}, which is an improvement of TransR for large-scale networks. Recently, the RotatE method has been developed \cite{Sun2019}. RotatE is a knowledge graph embedding method that can model and infer various relation patterns such as symmetry, inversion, and composition.
\end{itemize}

\subsection{Deep learning methods}
\hfill\\

In recent years, the use of deep learning for data analysis has increased steadily, and network analysis, including network embedding, is no exception. The success of deep learning methods can be explained by their ability to capture complex features and non-linearity among input variables. We define three major classes of deep learning embedding methods: the conventional neural networks based methods, the graph neural networks based methods, and the graph generative methods methods. All three of these classes of methods are based on different philosophies and mathematical formulations of deep learning.

\subsubsection{Conventional neural networks}
\hfill\\

The first network embedding methods based on deep learning methods use conventional deep learning techniques. We can cite the following class of deep learning architectures:

\begin{itemize}
    \item {\bf Autoencoder:} Autoencoders and their various variants been widely used for feature extraction. This capability has been utilized by several high-performing network embedding methods such as: DNGR (Deep Neural Networks for Learning Graph Representations) \cite{Cao2016}, SDNE (Structural deep network embedding) \cite{Wang2016}, VGAE (Variational graph auto-encoders) \cite{Kipf2016})
    \item {\bf Convolutional Neural Networks (CNN):} CNN methods have shown high performance \cite{Alzubaidi2021} for detecting the significant features, this ability has been used by various network embedding methods. DKRL (Description-Embodied Knowledge Representation Learning) \cite{Xie2016a}, PSCN (PATCHY- SAN) \cite{Niepert2016})
    \item {\bf Other Neural networks:} For instance, a Multi-Layer Perceptron (MLP) is used by the method PALE (Predicting Anchor Links via Embedding) \cite{Man2016}), and a Recurrent Neural Networks (RNN) is used by the method Deepcas \cite{Li2017a}). \\
\end{itemize}
        
\subsubsection{Graph Neural Networks (GNN)}
\hfill\\

Recently, an important class of deep learning methods for network embedding has been developed: Graph Neural Networks (GNNs) \cite{Defferrard2016}. GNNs generalise the notion of Convolutional Neural Networks (typically applied to image datasets, with an image seen as a lattice network of pixels) to arbitrary networks. GNNs encode high-dimensional information about each node neighbourhood into a dense vector embedding. GNNs algorithms can be divided into two main components. The encoder, which maps a node $v_{i}$ into a low-dimensional embedding vector $z_{i}$, based on the local neighbourhood and the attributes of the node, and a decoder, which extracts user-specified predictions from the embedding vector. This kind of method is suitable for end-to-end learning and offers state-of-the-art performance \cite{Defferrard2016, Zitnik2018}. GNNs and their application to network embedding can be divided into different classes of methods.

\begin{itemize}
    \item {\bf Graph Convolutional Networks (GCNs):} Graph Convolutional Networks (GCNs) is the generalisation of Convolutional Neural Networks (CNNs) to graphs \cite{Kipf2017}. The basic idea behind CNN is to apply convolutional operations during the learning process to capture local properties of the input data, recognising identical features regardless of the spatial locations. Several similar successful approaches have been developed, and we can mention Chebyshev Networks \cite{Defferrard2016} and SAGE \cite{Hamilton2017a}.
    \item {\bf Graph Attention Network :} A well-known shortcoming of the graph convolutions procedure is that they consider every node neighbour as having the same importance. Graph Attention Networks (GAT) are neural networks architectures that leverage masked self-attentional layers to address this shortcoming \cite{Velickovic2017, Xu2017, AbuElHaija2018}.
    \item {\bf Other Graph Neural Networks:} Several other embedding methods based on alternative architectures of GNNs exist \cite{Xu2018, Ying2018, Zhang2019c}. The interested reader can refer to the review of Zhou et al.~on GNNs for more details \cite{Zhou2020a}. \\
\end{itemize}
    
\subsubsection{Graph generative methods}
\hfill\\

Graph generative methods are other deep learning methods mostly known for Generative Adversarial Networks (GANs) \cite{Goodfellow2014}. The principle of GANs is based on two components: a generator, and a discriminator. The idea of GANs is to train a generator until it is efficient enough to mislead the discriminator. The discriminator is misled when it cannot discriminate real data from the data generated by the generator. Based on this idea, several embedding methods appeared, including GraphGAN \cite{Wang2018}, Adversarial Network Embedding  (ANE) \cite{Dai2018}, and ProGAN \cite{Gao2019}. In the case of GraphGAN, for a given vertex, the generator tries to fit its underlying true connectivity distribution over all other vertices and produces `fake' samples to fool the discriminator, while the discriminator tries to detect whether the sampled vertex is from the ground truth or generated by the generative model. An alternative method to GAN is the Restricted Boltzmann Machine \cite{McClelland1987}, which inspired different embedding methods (pRBM \cite{Wang2016a}).

\subsection{Higher-order network methods}
\hfill\\

We have seen in section 3.A that shallow embedding methods use pairwise similarity functions. This choice is imposed by the structure of the graphs, which by definition connect nodes by pairwise interactions. However, generalisations of graphs that can encode higher-order interactions, such as hypergraphs and simplicial complexes, are increasingly being studied \cite{bianconi2021higher,battiston2020networks, battiston2021physics,torres2021and}. Hypergraphs encode arbitrary relations between any number of nodes, that is, edges are generalised to hyperedges which can contain any number of nodes, not just two. Simplicial complexes generalise graphs by allowing triangles, tetrahedrons, and higher-dimensional `cliques' to be represented, and are closely related to Topology, particularly Topological Data Analysis \cite{Salnikov2018, zomorodian2012topological}. Note that simplicial complexes are a type of hypergraphs, so, at least in principle, hypergraph methods also apply to simplicial complexes. Recently, network embedding methods have been extended to consider these new types of higher-order networks. 
In particular, some of the existing network embedding methods described in sections 3.A-B have been extended to hypergraphs. Methods capable of embedding hypergraphs include shallow embedding methods: Learning hypergraph-regularized \cite{Huang2015} (matrix factorisation method), Spectral hypergraph embedding \cite{Zhou2006} (random walk method), LBSN2Vec++ \cite{Yang2020a} and HyperEdge-Based Embedding \cite{Gui2016, Gui2017} (optimisation methods). There are also methods derived from deep learning processes: Deep hypergraph network embedding \cite{Tu2018} (autoencoder), Hypergraph neural networks\cite{Feng2019} (GNN), Hypergraph attention embedding \cite{Bai2021} (GAT). As explained above, simplicial complexes are a special type of hypergraphs which are amenable to be treated by powerful tools based on algebraic topology \cite{papillon2023architectures}, including Topology Data Analysis (TDA) \cite{Salnikov2018, zomorodian2012topological}. The literature on simplicial complexes embedding and simplicial neural networks is rapidly growing \cite{papillon2023architectures}. It includes  simplicial and cell complex neural networks \cite{bodnar2021weisfeiler, bodnar2021weisfeiler2, ebli2020simplicial, chen2022bscnets, hajij2020cell, Schaub2021, giusti2022cell} and Geometric Laplacian eigenmaps embedding (Glee)\cite{torres2020glee}. Simple graphs can also be described as higher-order set-of-sets formed by node neighbourhoods, for which recently a new graph embedding has been proposed (HATS)\cite{meng2019hats}.

\subsection{Emerging methods}
    \hfill\\

Here we highlight some key emerging methods for network embeddings which deserve particular attention. Specificaly, we discuss key results in the rapidly growing literature focusing on network embeddings in non-Euclidean spaces, including hyperbolic and Lorentzian spaces. Moreover, we cover the very vibrant research activity on a new generation of neural networks using Magnetic and Connection Laplacians which provide powerful tools to treat directed networks and to improve the explainability of the algorithms. Finally, we discuss the important research direction aiming at comparing different embedding algorithms.

 \subsubsection{Network embedding in hyperbolic and Lorentzian spaces}
 \hfill\\

Embedding in hyperbolic spaces offers advantages according to different perspectives \cite{aste2005complex, kleinberg2007geographic}. Among the benefits of hyperbolic spaces, the most relevant one is probably the fact that hyperbolic spaces are natural spaces to embed trees having a number of nodes growing exponentially with the distance from the root, and in general to embed small-world networks.
We distinguish three major approaches to embedding networks in hyperbolic spaces: 
\begin{itemize}
    \item {\bf Embedding based on the complex hyperbolic network models \cite{krioukov2010hyperbolic, papadopoulos2012popularity} in the $\mathbb{H}^2$ plane.} According to the spatial hyperbolic network \cite{krioukov2010hyperbolic} and PSO \cite{papadopoulos2012popularity} models, the radial coordinate of the node embedding is determined by the degree of the nodes and the angular coordinate of the node embedding is determined by a similarity metric.
    The hyperbolic embedding can be used to formulate a greedy algorithm for network navigability \cite{boguna2010sustaining, kleinberg2007geographic}, to predict missing edges \cite{kitsak2020link}, and also to relate the clusters found along  the angular coordinates to network communities \cite{faqeeh2018characterizing, zuev2015emergence}. For a general review of this approach see Ref. \cite{boguna2021network}.
    The original embedding algorithm HyperMap \cite{papadopoulos2012popularity}, used to determine the angular coordinates of the nodes and revisited in \cite{nickel2017poincare}, maximizes the likelihood that the data is drawn from the model. Mercator \cite{garcia2019mercator} improves this algorithm by initializing the position of the nodes using Laplacian Eigenmaps and, for each optimisation step, the angular positions of the nodes is updated by choosing, among a set of several possible moves, the one that optimises the likelihood. The possible new moves are drawn from a Gaussian distribution centred on the mean angles among the neighbour nodes. This algorithm has computational complexity $O(n^2)$ on sparse networks. A fast and efficient alternative to this approach is provided by the {\em noncentered minimum curvilinear embedding} (ncMCE) \cite{muscoloni2017machine}  which provides a  machine learning pipeline including three main steps: (i) a pre-weighting procedure which identifies the network backbone; (ii) the extraction of  the matrix {\bf D} of nodes similarities (distances) measured on this network backbone (iii) a dimensionality reduction of the matrix {\bf D}.
    The ncMCE has been further extended in \cite{kovacs2021optimisation} to further reduce the loss calculated according to the PSO loglikelihood. Recently, an exact and rapid one-dimensional embedding algorithm \cite{patania2023exact} based on dynamic programming has been proposed. This algorithm can determine the angular coordinates of the hyperbolic embedding in $\mathbb{H}^2$ and, more generally, extract other types of one-dimensional embeddings. Finally, the hyperbolic embedding of directed networks has been addressed in Ref. \cite{kovacs2023model}. 
    \item{\bf Embedding of hierarchical data.}
    Hyperbolic spaces allow the embedding of hierarchical data and in particular trees without distortion in spaces of low dimension \cite{sarkar2012low} while the same data would require a high dimensional Euclidean embedding if low distortion is desired. This fundamental property of hyperbolic spaces has been exploited in \cite{sala2018representation} to first embed a network in a tree and then embed the tree into hyperbolic spaces achieving a fast and reliable hyperbolic embedding. This approach has been extended to knowledge graphs in Refs. \cite{chami2020low} while in \cite{chami2019hyperbolic} Hyperbolic Graph Convolutional Neural Networks have been proposed to learn the hyperbolic embedding.  Note that embedding  a network into another network (which might not in general be a tree) is a generalised embedding problem tackled also in Ref.\cite{fernandez2019flexible}.
    \item {\bf Filtering of networks generating the  Triangulated Maximally Filtered Graph (TMFG) \cite{massara2016network}.} The TMFG  generalises the Minimal Spanning Tree and has the topology of a `fat tree' formed by $d$ connected $(d+1)$-cliques ($d$-simplices). The structure of TMFG reduces to the structure of the model {\em Network Geometry with Flavor} \cite{bianconi2016network} with natural hyperbolic embedding in $\mathbb{H}^d$ \cite{bianconi2017emergent}.
    Therefore TMFGs have a hyperbolic geometry while the previously proposed maximally filtered planar graphs  \cite{tumminello2005tool} have a natural $\mathbb{R}^2$ embedding.
\end{itemize}
  Finally, we point out also that network embeddings in Lorentzian (Minkowskian) spaces have been proposed \cite{clough2017embedding} and applied to the study of directed acyclic networks as citation networks. This embedding can be used for  paper recommendation, identifying missing citations, and fitting citation models.
  
\subsubsection{Network embeddings using Magnetic and Connection Laplacians}
\hfill\\

Despite many networks being directed, machine learning methods are typically developed for undirected networks. One challenge in directed networks is that they are naturally encoded by asymmetric adjacency matrices with a complex spectrum while machine learning techniques usually require a loss function that is real and positive definite. In order to address this challenge, the use of magnetic Laplacians is attracting growing attention. Magnetic Laplacians are Hermitian matrices (hence having a real and non-negative spectrum) that encode the direction of edges through a complex phase. Using a magnetic Laplacian, it is possible to formulate Eigenmap embeddings \cite{fanuel2018magnetic, gong2021directed, gong2023generative} that can detect non-trivial periodic structure in the network such as three or more communities whose connection pattern is cyclic. Additionally, in Ref. \cite{zhang2021magnet}, the magnetic Laplacian is also used to propose {\em Magnet}, a novel and efficient neural network for directed networks. This vibrant research activity on the magnetic Laplacian is indicative of the recent interest in network structure with complex weights \cite{bottcher2022complex}. 
The Magnetic Laplacian can be considered a special case of the Connection Laplacian, which can be used for Vector Diffusion Maps \cite{singer2012vector}. Moreover, the Connection Laplacian is used for  formulating  Sheaf Neural Networks \cite{bodnar2022neural,barbero2022sheaf},  which are a new generation of neural networks obtaining excellent performance on several Machine Learning tasks.
Finally, the  non-backtracking matrix that identifies non-backtracking cycles and efficiently detects the network communities \cite{krzakala2013spectral}, has been recently proposed for embedding oriented edges (Non-backtracking embedding dimensions: NBED) \cite{torres2019non}.

\subsubsection{Comparison of different algorithms}
 \hfill\\

An important question is how to choose among the different network embedding algorithms and how to select their hyperparameters. For instance, the selection of the embedding dimension is a crucial hyperparameter for network embedding algorithms. Ref. \cite{gu2021principled} provides a method to determine the value of the embedding dimension that constitutes the best trade-off between favouring low dimensions and requiring the generation of a network representation close to the best representation that the chosen algorithm can achieve. The comparison between different embedding algorithms \cite{zhang2021systematic} and the investigation of network embedding algorithms as opposed to alternative inference approaches such as community detection \cite{tandon2021community} is attracting increasing attention and can guide the choice of the most suitable algorithm. Finally, we note that a unifying approach of node embeddings and structural graph representation has been recently proposed \cite{srinivasan2019equivalence}.

\section{Applications}
\hfill

A wide variety of applications exist for network embedding methods. Some methods have been designed focusing on specific task(s) but others admit versatile applications. In the following sections, we will present the most common applications of network embedding methods as well as some emergent applications offering promising results. We associate each application with some widely used embedding methods in order to give the interesting reader a guideline for applying network embedding in different tasks (Fig.~3).

\begin{table}
    \centering
    \setlength{\tabcolsep}{12pt}
    \renewcommand{\arraystretch}{1.5}
    \begin{tabular}{c|c}
        {\bf Operator}    & {\bf Definition} \\ \hline
        Average           & $\frac{z_{i}(k) + z_{j}(k)}{2}$ \\ \hline
        Hadamard          & $z_{i}(k) \odot z_{j}(k)$ \\ \hline
        Weighted-L1       & $| z_{i}(k) - z_{j}(k) |$ \\ \hline
        Weighted-L2       & $| z_{i}(k) - z_{j}(k) |^{2}$ \\ \hline
        Cosine            & $\frac{z_{i}(k) \cdot z_{j}(k)}{\lVert z_{i} \rVert \lVert z_{j} \rVert}$ 
    \end{tabular}
    \caption{Binary operators to compute infer edges between pairs of embedding vectors. The variable $z_{i}$ defines the embedding vector associated with the node $v_{i}$, and $z_{i}(k)$ defines the $k\text{-th}$ element of the embedding vector$z_{i}$.}
\end{table}

\subsection{Classical applications of network embedding}
\hfill\\


\begin{itemize}
    \item {\bf Node classification:} The aim is to predict labels for unlabelled nodes based on the information learned from the labelled nodes. Network embedding methods embed nodes into vectors which can be used in an unsupervised setting. In this case, the nodes associated with similar node embedding vectors will have similar labels. In a supervised setting, the classifier is trained with the vectors associated with the labelled nodes. The classifier is then applied to predict the labels of unlabelled nodes.
    \item {\bf Link prediction:} The aim is to infer new interactions between pairs of nodes in a network.
    The similarity between the nodes encodes the propensity of the nodes to be linked. The similarity can be computed, for instance, with an inner product or a cosine similarity between each pair of node embedding vectors. In the embedding space, several operators exist to infer edges between pairs of embedding vectors \cite{Grover2016}. These operators can be, for instance, binary operators (Table 1), or, heuristic scores (Table 2).
    \item {\bf Node clustering/Community detection:} The aim is to determine a partition of the network such that the nodes belonging to a given cluster are more similar to each other than to the nodes belonging to other clusters. In practice, any classic clustering method can be directly applied in the latent space to cluster the nodes. K-means \cite{Macqueen1967} is often used for this purpose. However, it is useful to notice that some embedding methods are designed specifically for this task \cite{Ahmed2013}.
    \item {\bf Network reconstruction:} The aim here is to reconstruct the whole network based on the learned embedding representation. Let us define $n$ as the total number of nodes in the network. The reconstruction imposes $n(n-1)/2$ evaluations to test each potential edge, and each evaluation is equivalent to a link prediction.
    \item {\bf Visualisation:} Network embedding methods, as other reduction dimension methods, can be used to visualise high-dimensional data in a lower-dimensional space. 
    We expect similar nodes to be close to each other in the visualisation. However, network embedding methods that were not designed for this specific task show poor results when directly projected into a two-dimensional embedding space \cite{Tang2015, Zhang2020}. To bypass this shortcoming, the visualisation of the result of network embedding methods is frequently projected into a $d$-dimensional  embedding space ($2 < d \ll n $), and then into a two-dimensional space. The two-dimensional space is obtained with a dimension reduction method suitable for visualisation, like Principle Component Analysis (PCA) \cite{Wold1987, Jolliffe2016}, t-distributed Stochastic Neighbor Embedding (t-SNE) \cite{Maaten2008}, or Uniform Manifold Approximation and Projection (UMAP) \cite{McInnes2018}. Recently, an extension of PCA for data lying in hyperbolic spaces has been developed \cite{Chami2021}. \\
    Finally, visualisation is a powerful tool to investigate the results obtained by embedding methods. The interpretability of results can be enhanced through visualisation, and a recent method has been proposed to address questions related to bias and fairness in results \cite{Rissaki2022}.
\end{itemize}

\begin{table}
    \centering
    \setlength{\tabcolsep}{12pt}
    \renewcommand{\arraystretch}{1.8}
    \begin{tabular}{c|c}
        {\bf Score}             & {\bf Definition} \\ \hline
        Common Neighbors        & $| \mathcal{N}(v_{i}) \cap \mathcal{N}(v_{j}) |$ \\ \hline
        Jaccard’s Coefficient   & $| \frac{\mathcal{N}(v_{i}) \cap \mathcal{N}(v_{j})}{\mathcal{N}(v_{i}) \cup \mathcal{N}(v_{j})} |$         \\ \hline
        Adamic-Adar Score       & $\sum_{t \in \mathcal{N}(v_{i}) \cap \mathcal{N}(v_{j})} \frac{1}{\ln{|\mathcal{N}(v_{t})|}}$ \\ \hline
        Preferential Attachment & $| \mathcal{N}(v_{i}) |  | \mathcal{N}(v_{j})|$    
    \end{tabular}
    \caption{Heuristic scores are used to predict edges between pairs of nodes in the direct space. The variable $\mathcal{N}(v_{i})$ defines the neighbor set of nodes associated with the node $v_{i}$.}
\end{table}

\begin{figure*}[ht]
	\captionsetup{justification=centering}
    \centering
	\includegraphics[width=16cm,height=16cm,keepaspectratio]{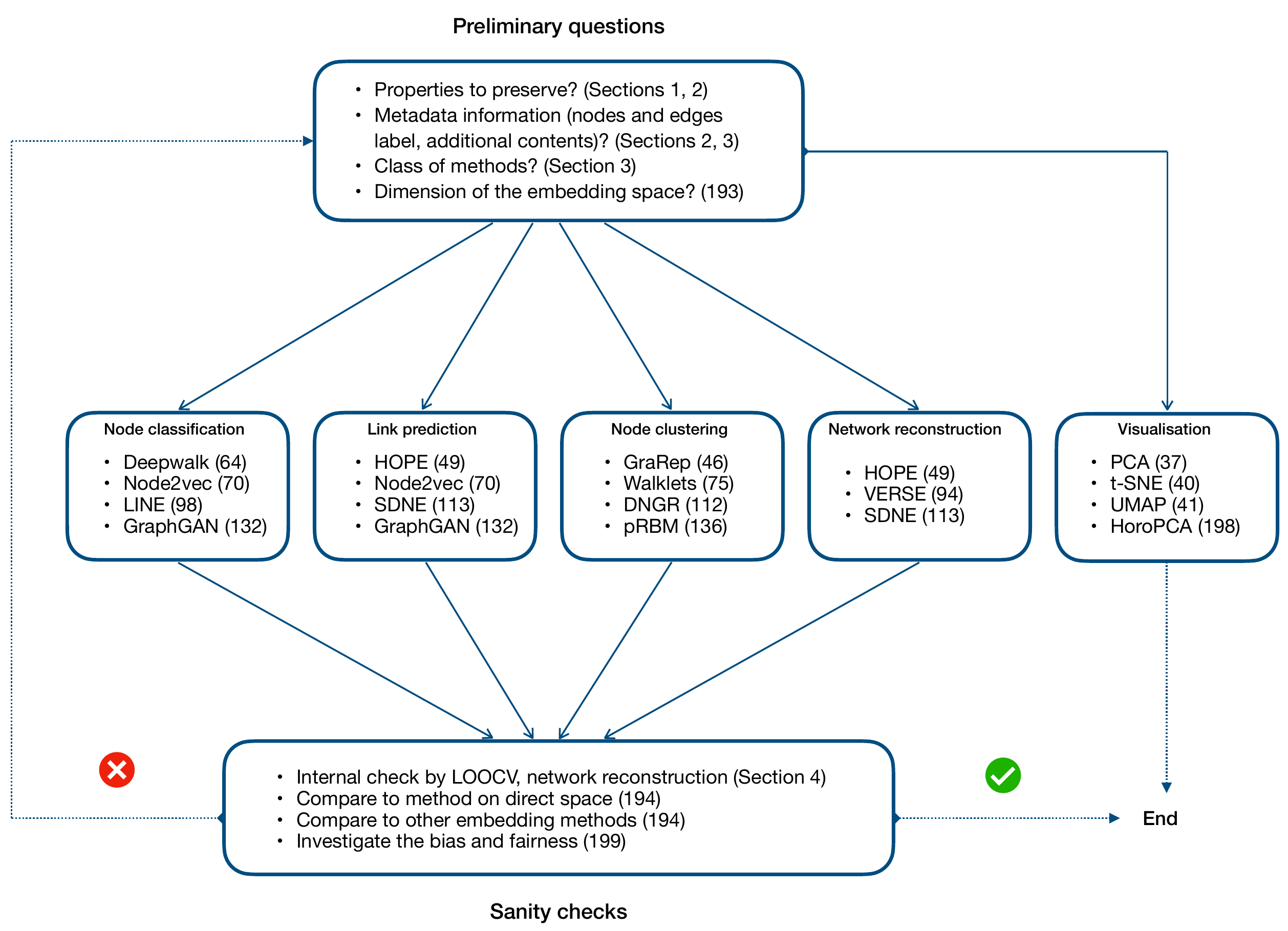}
    \caption{Workflow for choosing a network embedding method. The application of network embedding starts with some preliminary questions (top box). Depending on the answers, and the task to be performed, different network embedding methods can be applied. Here, we list the most common methods associated with the most common network embedding tasks. Once the embedding representation has been obtained, the workflow could further perform some `sanity checks' to measure the efficiency of the network embedding setup (bottom box). Depending on the results of these sanity checks, we can either stop the development or go back to the preliminary questions in order to improve the workflow. These improvements are usually based on adding complementary information, tuning the parameters, or redefining the properties to be preserved by the embedding representation.}
\end{figure*}

\subsection{Emerging applications}
\hfill\\

\begin{itemize}
    \item {\bf Network compression and coarsening:} The aim of network compression is to convert a large network into a smaller network containing a reduced number of nodes and edges. This compression is expected to store the network more efficiently and to allow running the network algorithms faster. 
    Network coarsening is often used as a preliminary step in the network embedding process to produce a compression by collapsing pairs of nodes and edges with appropriate criteria. One example is network compression using symmetries. Real-world networks have a large number of symmetries \cite{macarthur2008symmetry, macarthur2009spectral} and this can be exploited in practice for compression or coarsening, as well as for speeding up calculations \cite{sanchez2020exploiting, wang2012symmetry}.

    \item {\bf Network classification:} The aim is to associate a label to a whole network. Network classification can easily be applied in the context of whole network embedding. A wide range of applications has been proposed, such as classifying molecular networks according to their properties \cite{Dai2016, Duvenaud2015, Kearnes2016, Niepert2016}, predicting therapeutic effects of candidate drugs based on molecular networks \cite{Kearnes2016}, or classifying images that have been converted into networks representation \cite{Bruna2014}.
    
    \item {\bf Applications to Knowledge graphs:} Let us consider a knowledge graph defined by a set of triples $(u,r,v) \in V\times R\times V$. There are three main applications of embedding for knowledge graphs. Link prediction is used to infer the interaction between $u$ and $v$. Triplet classification, which is a standard binary classification task, determines if a given triplet $(u,r,v)$ is correct. And finally, knowledge completion aims to determine the missing element in a triplet where only two of the pieces of information are known \cite{Li2020, Feng2016}.

    \item {\bf Illustration of network embedding with biological applications:} Network embedding is an active research topic in bioinformatics \cite{Li2022}, and all the classical applications previously mentioned also flourish in the context of biological networks. Some applications emerge as particularly relevant to network biology. We describe here some of them and the interested reader can refer to \cite{Nelson2019, Li2022} for detailed reviews. A first important application in biology is related to network alignment, which aims to find correspondences between nodes in different networks. This can be useful to reveal similar subnetworks, and thereby biological processes, in different species, by aligning their protein-protein interaction networks \cite{Fan2018, Heimann2018}. Another important application pertains to network denoising, which consists of projecting a graph into an embedding space to reduce noise by preserving the most relevant properties of the original network. For instance, high-order structures of the original networks can be preserved by diffusion processes. Network embedding methods can also be used to predict the functions of proteins \cite{Zitnik2017}, or to detect modules in chromosome conformation networks \cite{Nelson2019}. Laplacian Eigenmaps have also been used to map the niche space of  bacterial ecosystems from genomic information \cite{thilo2020mapping}. In genomics, UMAP embedding has been shown to be useful to identify  overlooked sub-populations and characterise fine-scale relationships between geography, genotypes, and phenotypes in a given population \cite{diaz2019umap}. Moreover, the prediction of edges (also known as associations in the biological context) is a common task in biological network embedding. For instance, a recent method named GCN-MF \cite{Han2019} used Graph Convolutional Networks and matrix factorisation to predict gene-disease associations. Another method, named NEDTP \cite{An2021}, used a multiplex network embedding based on an optimisation method to predict drug-target associations.
    Finally, knowledge graphs are also used in biomedical contexts. For example, Electronic Health Record (EHR) can be represented as a knowledge graph and embedded jointly with other networks integrating proteins, diseases, or drug information, to predict patient outcomes \cite{Rotmensch2017, Wu2019}.
\end{itemize}

\hfill
\section*{Conclusion}
\hfill

Network embedding methods are powerful tools for transforming high-dimensional networks into low-dimensional vector representations that can be used for visualization and a wide range of downstream analyses, such as network inference, link prediction, node classification, and community detection. The resulting low-dimensional representation also enables the use of deep-learning analysis, which is not possible directly on the original network as it is inherently a combinatorial object. \\
Moreover, the latent space generated through the embedding process can highlight relevant features of the original network and filter out the noise inherent in the dataset used to construct the network. This helps to reveal the underlying structure and organization of the network, making it easier to analyze and interpret its properties. \\
Despite the numerous advantages of network embedding methods, there are still significant questions surrounding their use, particularly with regard to selecting appropriate methods and assessing their properties, interpreting the results, and ensuring adaptability across a range of contexts. These questions are especially critical as datasets become more complex and require more sophisticated tools to achieve the desired level of adaptability and interpretability, particularly in the biological contexts, where datasets may be particularly noisy, leading to errors in the learning process. \\
In recent years, there has been a substantial increase in the number of network embedding methods, making it challenging to stay up-to-date in this rapidly evolving field. Therefore, it is crucial to have access to a comprehensive review of the state-of-the-art methods. Our network embedding review provides not only a summary of the current state-of-the-art, but also lays the foundation for future developments by presenting a flexible yet rigorous mathematical perspective-based taxonomy of embedding methods. Despite the many advancements in the field, challenges remain regarding the selection and interpretability of these methods. To address these issues, we offer a set of guidelines for practical applications that take into account these considerations and are aligned with the latest developments in the field.

\acknow{We acknowledge interesting discussions with Filippo Radicchi and  funding from the  Roche-Turing Partnership (A.Bap., R.S.-G.,G.B.) and the «Investissements d'Avenir» French Government program managed by the French National Research Agency (ANR-16-CONV-0001 and ANR-21-CE45-0001-01)(A.Bau.). We also acknowledge Galadriel Brière for her proofreading.}
\showacknow{} 

\bibliography{review_embedding}

\end{document}